\documentclass[conference]{IEEEtran}
\IEEEoverridecommandlockouts
\usepackage{cite}
\usepackage{amsmath,amssymb,amsfonts}
\usepackage{algorithmic}
\usepackage{graphicx}
\usepackage{textcomp}
\usepackage{xcolor}
\usepackage{multirow}
\usepackage{float}
\usepackage{tikz}
\usepackage[caption=false,font=footnotesize]{subfig}

\def\BibTeX{{\rm B\kern-.05em{\sc i\kern-.025em b}\kern-.08em
    T\kern-.1667em\lower.7ex\hbox{E}\kern-.125emX}}
\begin{document}

\newcommand{\ie}{\textit{i.e.}}
\newcommand{\eg}{\textit{e.g.}}
\newcommand{\fig}[1]{Fig.~\ref{#1}}
\newcommand{\tab}[1]{Table~\ref{#1}}
\newcommand{\lib}[1]{\textsc{#1}}

\title{Early Exiting U-Net for Efficient Processing on UAVs: A Case Study in Environmental Monitoring}

\author{\IEEEauthorblockN{Luca Sartori Boni, Mohamed Moursi, Norbert Wehn, Bilal Hammoud}
\IEEEauthorblockA{\textit{Microelectronic Systems Design (EMS), RPTU, Kaiserslautern-Landau, Germany} \\
Email: luca.boni@edu.rptu.de, mmoursi@rptu.de, norbert.wehn@rptu.de, bilal.hammoud@rptu.de}
}

\maketitle

\newcommand\submittednotice{%
\begin{tikzpicture}[remember picture,overlay]
\node[anchor=south,yshift=10pt] at (current page.south) {\fbox{\parbox{\dimexpr0.65\textwidth-\fboxsep-\fboxrule\relax}{\submittedtext}}};
\end{tikzpicture}%
}

\newcommand\copyrighttext{%
  \footnotesize \textcopyright \the\year{} IEEE. Personal use of this material is permitted. Permission from IEEE must be obtained for all other uses, including reprinting/republishing this material for advertising or promotional purposes, collecting new collected works for resale or redistribution to servers or lists, or reuse of any copyrighted component of this work in other works.}

\newcommand\copyrightnotice{%
\begin{tikzpicture}[remember picture,overlay]
\node[anchor=south,yshift=10pt] at (current page.south) {\fbox{\parbox{\dimexpr0.75\textwidth-\fboxsep-\fboxrule\relax}{\copyrighttext}}};
\end{tikzpicture}%
}

\copyrightnotice

\begin{abstract}
Oil spills represent a severe threat, making early-stage thickness estimation crucial for guiding remediation efforts.
Unmanned Aerial Vehicles (UAVs) are an attractive platform for environmental monitoring.
However, due to their limited computation and power budgets, real-time onboard processing requires optimized algorithms or lightweight machine learning models.
While the standard U-Net architecture is often too large for constrained UAV hardware, the compressed Tiny U-Net variant fits on FPGA platforms and achieves competitive estimation performance (0.79 in the metric Intersection over Union, or IoU). Despite this success, Tiny U-Net processes every radar image through the complete inference pipeline, resulting in unnecessary computation for simple cases.
To address this inefficiency, we integrate an early exit feature into the Tiny U-Net architecture. We introduce an early exit branch that returns an early prediction when a compact confidence score exceeds a tunable threshold, bypassing deeper layers for high-confidence evaluations. Our experiments demonstrate that this design achieves comparable IoU to the full baseline model. Crucially, the technique is shown to reduce the average number of multiplications by up to 42\% for an aggressive threshold, reducing the dynamic power consumption.
Choosing a threshold that ensures extreme confidence reduces the complexity-reduction gains for an improved IoU. This early exit approach substantially improves computational efficiency in Tiny U-Net, enabling more practical deployment in UAV-based environmental monitoring systems.
\end{abstract}


\section{Introduction}

Oil spills represent one of the most severe threats to marine ecosystems, coastal economies, and offshore activities; their ecological and economic consequences have been widely documented. Early-stage thickness estimation is crucial for guiding containment and remediation efforts. Accurate, on-site thickness information improves decision-making for response teams and reduces the environmental and economic impacts of spills~\cite{i1},~\cite{i2}.

Unmanned aerial vehicles (UAVs) have become increasingly attractive for environmental monitoring due to their low cost, high spatial resolution, and rapid deployability \cite{i3}. UAV-mounted sensors have been deployed for oil detection tasks using visible \cite{i4}, infrared \cite{i5}, and hyperspectral modalities \cite{i6}, but these sensors can be limited by daylight or weather conditions.
Radar-based sensors provide a robust alternative that can operate across a wider set of conditions and are well suited for on-site thickness estimation \cite{dote}.
While onboard processing enables rapid responses, real-time onboard processing on battery-powered UAVs is constrained by limited computation, memory, and power budgets. Thus, lightweight processing models and hardware-aware designs are required to enable on-edge inference \cite{i7}, \cite{i8}.

The convolutional U-Net architecture \cite{unet} has become a standard for dense per-pixel estimation tasks (segmentation and thickness mapping) because of its encoder–decoder structure with skip connections. However, the full U-Net is often too large for direct deployment on constrained UAV hardware. To bridge this gap, a compressed Tiny U-Net variant was recently proposed and evaluated for UAV-based oil-spill thickness estimation \cite{tu}. It reduces the number of convolutional blocks and channels to fit on FPGA platforms, while still achieving competitive performance and low runtime power consumption in hardware tests.

While Tiny U-Net demonstrates that aggressive architectural compression and hardware-aware optimizations can enable on-edge thickness estimation \cite{tu}, the model still applies the same full-depth inference pipeline to every radar image, regardless of the relative difficulty of each scene. This means that simpler inputs still incur the full computational cost of the network even when a shallower representation could yield sufficiently confident predictions. Early exit (EE) deep neural networks (DNNs) address this inefficiency by adding auxiliary classifiers at intermediate layers: when an intermediate classifier reaches high confidence, the network can terminate inference early. This saves computation and reduces latency (a valuable property for real-time, energy-constrained UAV systems). EE DNNs have been used in mobile and edge scenarios to trade off accuracy and cost and to improve robustness in training and inference \cite{ee1}, \cite{ee2}.

In this work, we combine these two lines of research and tackle a real constraint in UAV-based marine pollution monitoring: limited computational resources and power budgets. Building on the Tiny U-Net design and on the same synthetic radar dataset and environmental model used in \cite{tu} (so results are directly comparable), we introduce an early exit branch between the encoder and bottleneck stages of Tiny U-Net. The EE branch produces an auxiliary thickness map and a compact confidence score per scene; when the score exceeds a tunable threshold, the model returns the early prediction; otherwise, it continues through the full Tiny U-Net pipeline. We evaluate the performance–efficiency tradeoffs of this design by sweeping the confidence threshold and reporting IoU, early exit rates, and average computational cost (measured as multiplication counts).
Our experiments show that the proposed Tiny U-Net with early exit achieves comparable IoU to the baseline while reducing the average number of multiplications by 42\% for an aggressive threshold, with only 2\% IoU degradation at practical threshold values.
\section{Previous Work}

\subsection{Environmental Model}

We adopt the same marine environment model used in \cite{dote}. The model consists of floating thick slicks at an early stage of development, from a few hours up to two days after the spill, before emulsification begins, with thicknesses ranging between 1 and 10~mm. The roughness of the ocean surface is represented by the surface root-mean-square height, which is influenced by wind speeds measured 10~m above the sea surface, varying from 2 to 8~m/s. The oil is assumed to have a dielectric constant of 3, while the water is modeled with a temperature of 20~°C and a salinity of 35~ppt.

\subsection{Monitoring System Model}

To estimate the spatial thickness distribution of oil spills on-site, we base our experiments on a modeled UAV system concept equipped with a wide-band radar sensor operating in the 4–12~GHz range, with 1~GHz increments across C- and X-bands. The radar is simulated as a nadir-looking system, where the incident electromagnetic waves are normal to the sea surface, allowing the full specular backscattering to be captured \cite{ev}. The onboard processing unit is represented by an FPGA-based platform, reflecting realistic power and latency constraints.

\subsection{Tiny U-Net Model}

The Tiny U-Net model \cite{tu} is a smaller version of the original U-Net model \cite{unet}. It uses only two convolution blocks (versus four in standard U-Net), with channels reduced to one quarter. Specifically, the first block has 16 channels, the second has 32, and the bottleneck layer has 64.

The input is a 3-D matrix with dimensions $\mathrm{W} \times \mathrm{L} \times \mathrm{R}$. $\mathrm{W}$ and $\mathrm{L}$ represent the geometrical width and length of the scanned environment, respectively, while $\mathrm{R}$ represents the number of radar channels. The output is a thickness map of the potential oil slicks, with dimensions $\mathrm{W} \times \mathrm{L} \times \mathrm{C}$. $\mathrm{C}$ represents the number of thickness classes, which in our case are 11 classes in the range \{0:1:10\}~mm, with 0~mm corresponding to the detection of clean water surface \cite{tu}.
\section{Early Exit Deep Neural Networks}

Conventional DNNs process all inputs through the full stack of layers, regardless of input complexity \cite{ee1}. EE DNNs introduce auxiliary classifiers, or `exit branches', at selected intermediate layers. When an input reaches such a branch, the model evaluates prediction confidence; if the confidence exceeds a predefined threshold, inference terminates early, thereby reducing computation and latency \cite{ee2}.

This design is particularly advantageous in resource-constrained or real-time environments, such as mobile devices and edge computing, where minimizing energy consumption and response time is critical. In addition to computational savings, early exit architectures help mitigate issues such as vanishing gradients, overfitting, and overthinking by providing additional gradient signals and regularization \cite{ee1}.
\section{Tiny U-Net Model with Early Exit Branch}

\subsection{Introduction of the Early Exit Branch}

\begin{figure*}[th]
\centerline{\includegraphics[width=\linewidth]{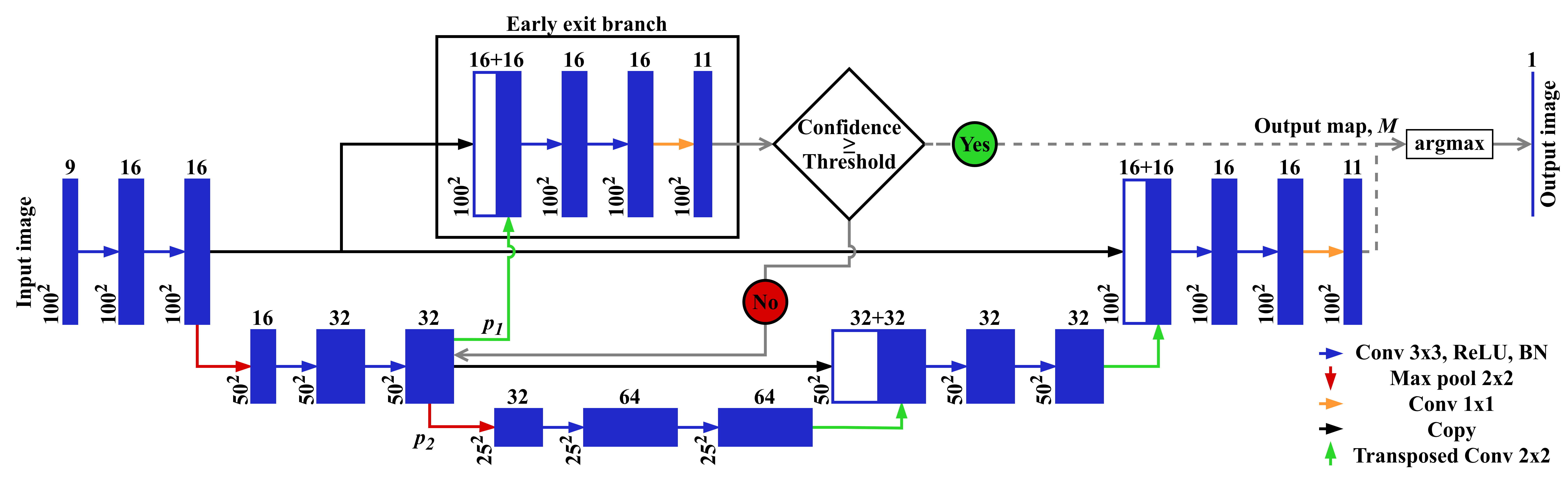}}
\caption{Architecture of the Tiny U-Net model with early exit branch. In this work, we use a resolution of $100\times100$, contrary to the $128\times128$ used in \cite{tu}.}
\label{model_architecture}
\end{figure*}

Rather than processing every radar image through the full stack of layers, we introduce an EE branch between the second convolution block and the bottleneck layer. In this configuration, the second convolution block serves as the bottleneck layer when inference exits early. \fig{model_architecture} shows the architecture of the model. After reaching the second convolution block, the data pass through the path $p_1$ to the early exit branch. Once the confidence of the output is calculated, the model decides whether to terminate computation or to return to the second layer and proceed through $p_2$, as in the original Tiny U-Net.

\subsection{Evaluating Confidence in the Early Exit Branch}

Given the output map $\boldsymbol{M}$ of the EE branch, with dimensions $\mathrm{W} \times \mathrm{L} \times \mathrm{C}$ (width, length, number of classes, respectively), its confidence $C_{\boldsymbol{M}}$, between 0 and 1, is determined as

\begin{equation}
C_{\boldsymbol{M}} = \dfrac{1}{\mathrm{W} \cdot \mathrm{L}}\sum_{i = 1}^{\mathrm{W}}\sum_{j=1}^{\mathrm{L}} \mathrm{max}(\mathrm{softmax}(\boldsymbol{M}_{i,j,:}))\text{.}
\end{equation}

This value is then compared to the confidence threshold $T$, defined by the use case. If $C_{\boldsymbol{M}} \geq T$, the model is confident enough in the output of the EE branch and can terminate the computation early. $T$ affects only inference, does not need to be defined during training, and can be adjusted at any time without requiring retraining of the model.

\subsection{Training}

To train the model, we used the same 555 oil spill scenarios from \cite{tu}: 504 for training and 51 for testing. We used categorical cross-entropy loss, and normalized all inputs to have zero mean and unit standard deviation. We trained the model for 25 epochs with stochastic gradient descent, saving the model on the best-performing epoch. We chose to use an exponential decay on the learning rate, with an initial value of 3$\cdot$10$^{-\text{4}}$ and a multiplicative factor of 0.98. We conducted training using the \lib{PyTorch} library using 32-bit floating-point, with 8-bit integer quantization-aware training provided by the \lib{Brevitas} library.
To count parameters and number of operations, we used the \lib{fvcore} library.

To train both paths, we perform two forward passes per training instance: one through the full path and one through the early exit branch. Then, the average loss between both inferences is used to back-propagate the errors and update the weights of both paths.

\subsection{Selection of the Confidence Threshold}

After training the model, the confidence threshold used for inference has to be chosen. To better find an optimal range, we evaluated the mean IoU against different confidence thresholds. This metric can be defined as

\begin{equation}
\text{Mean IoU} = \dfrac{1}{\mathrm{C}} \sum_{n=1}^{\mathrm{C}}\dfrac{\mathrm{TP}_n}{\mathrm{TP}_n+\mathrm{FP}_n+\mathrm{FN}_n} \text{,}
\end{equation}
\\ where $\mathrm{C}$ is the number of classes and $\mathrm{TP}$, $\mathrm{FP}$, and $\mathrm{FN}$ represent the true positive, false positive, and false negative pixels in the output image of the model, respectively.

We tested values for the confidence threshold between 0.850 and 0.990 with increments of 0.001.
\newcommand{\minitab}[2][l]{\begin{tabular}{#1}#2\end{tabular}}
\newcommand{\mr}[2]{\multirow{#1}{*}{\minitab[c]{#2}}}

\section{Results}

In this section, we evaluate the performance-efficiency tradeoffs of our Tiny U-Net with EE branch, by measuring IoU, early exit rates, and average number of multiplications across several values for $T$. We also benchmark our results to the original work in \cite{tu}.

\subsection{Early Exit or Not}

\begin{figure}[htb]
\centerline{\includegraphics[width=0.96\linewidth]{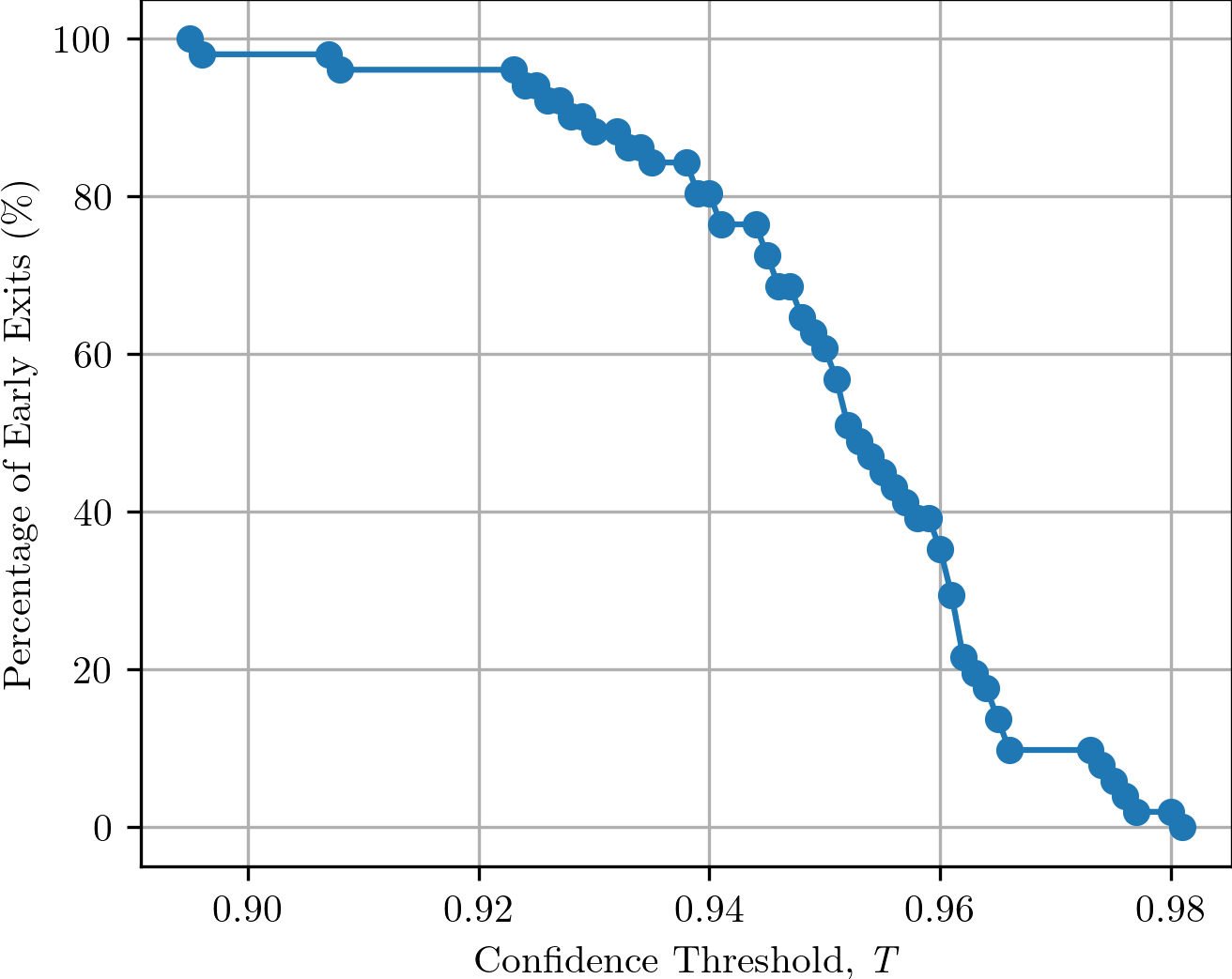}}
\caption{Percentage of samples that exited early versus confidence threshold.}
\label{ees}
\end{figure}

The added EE block increases model parameters slightly: 127\,622 versus 118\,395 in the original Tiny U-Net.
The confidence threshold directly affects the number of inferences that run through the full model and the ones that exit through an early branch. \fig{ees} shows the relation between the two; the lower the threshold, the fewer samples exit early. As a consequence, the model's performance and computational cost are influenced by $T$.

\begin{figure*}[t]
\centering
\begin{minipage}[t]{0.48\textwidth}
  \centering
  \includegraphics[width=\linewidth]{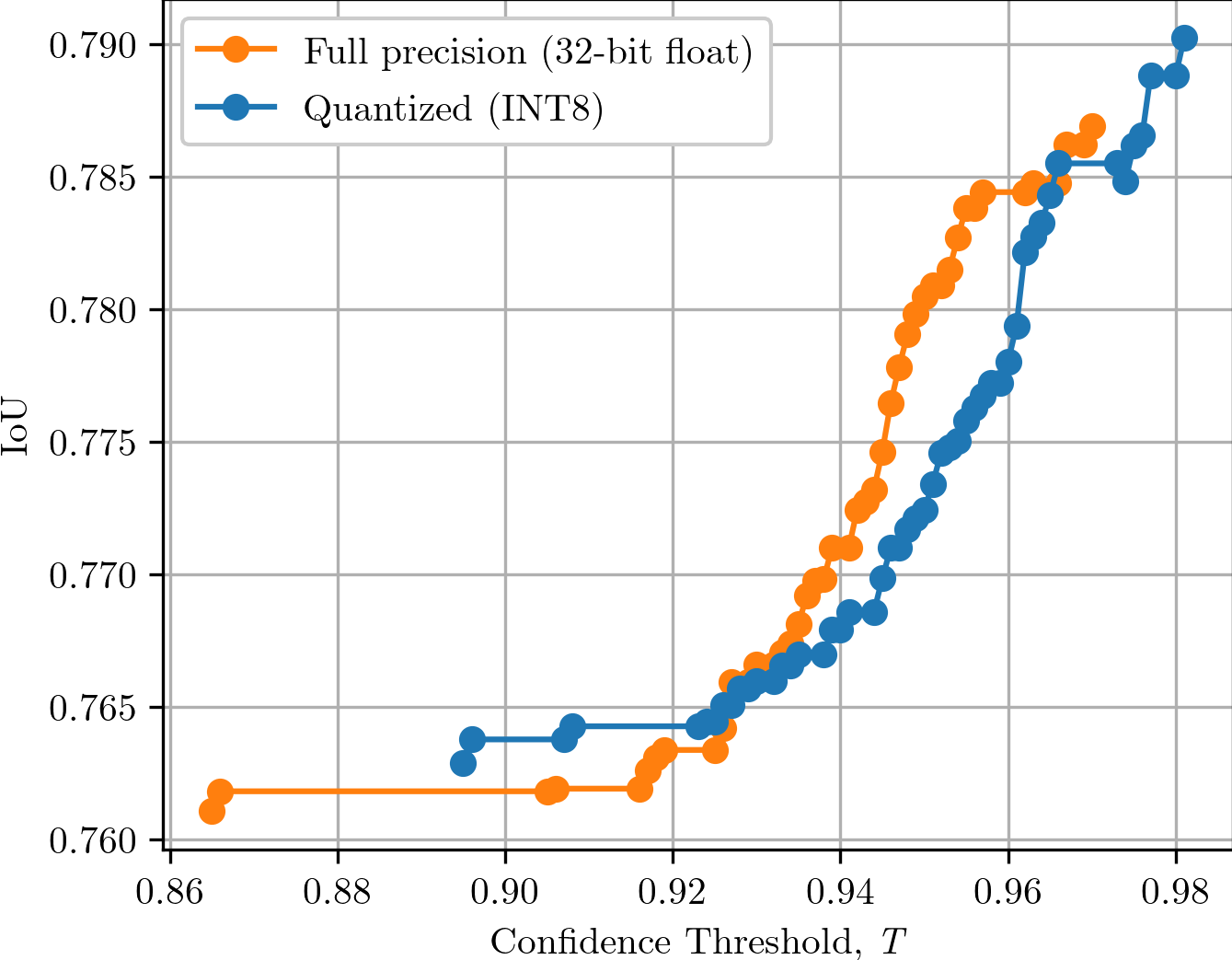}
  \caption{Average IoU over the test dataset versus confidence threshold.}
  \label{iou}
\end{minipage}
\hfill
\begin{minipage}[t]{0.47\textwidth}
  \centering
  \includegraphics[width=\linewidth]{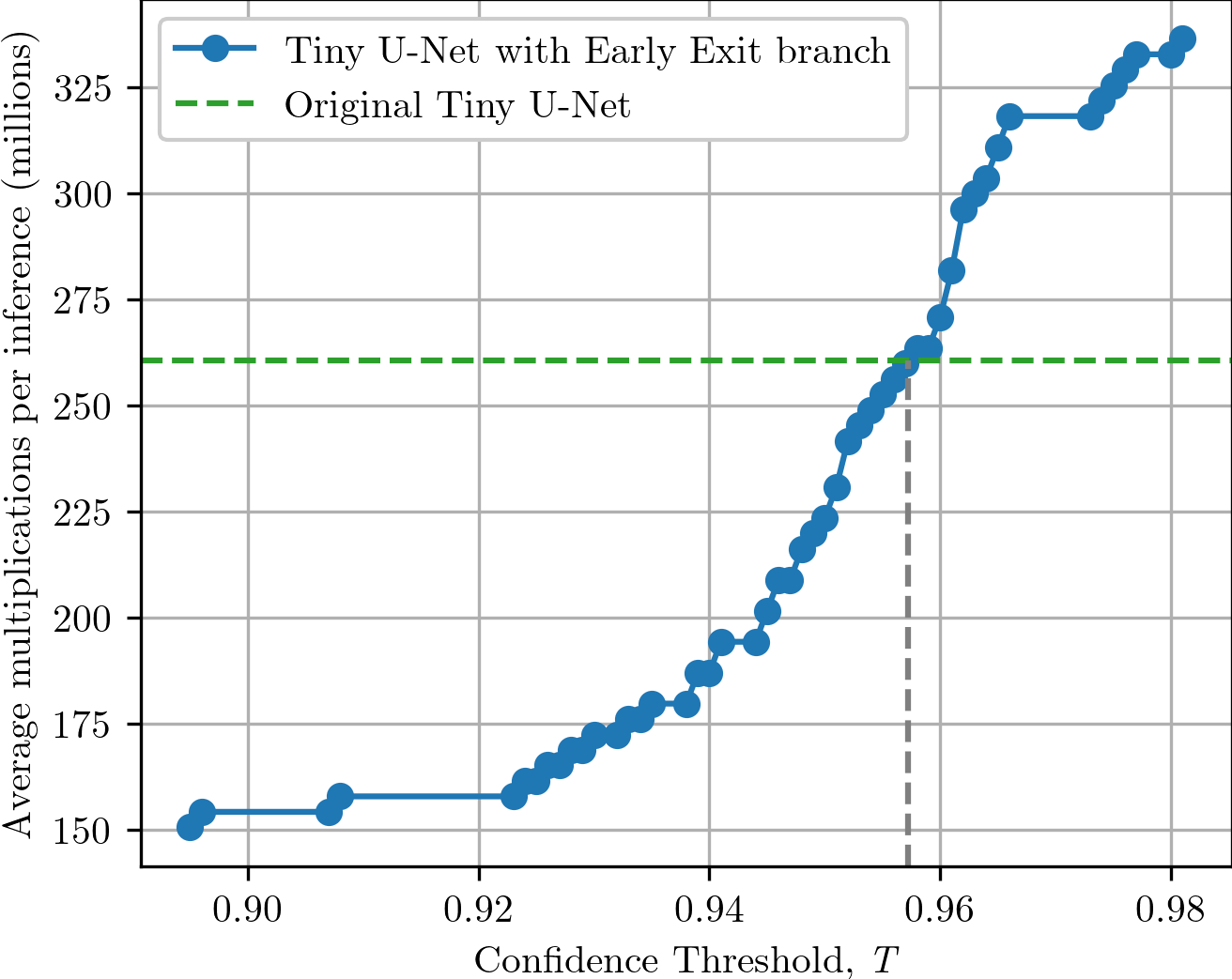}
  \caption{Average number of multiplications per inference over the test dataset versus confidence threshold.}
  \label{macs}
\end{minipage}
\end{figure*}

\subsection{Performance}

The original Tiny U-Net was able to achieve an IoU of 0.79 \cite{tu}. When all samples are forced to pass through the full network ($T \geq 0.981$), our model achieves similar performance. If we decrease the confidence threshold, the IoU also decreases, as can be seen in \fig{iou}, reaching 0.76 when all samples exit through the early branch.

Notably, the performance achieved by the model using only the EE branch surpasses that of a similar architecture explored in \cite{tu}: the model with one block and feature size 16 scored 0.70 in the IoU metric.

\subsection{Computational Cost}

When running in full precision, the Tiny U-Net performed 260.56 million multiplications. The model with the early exit branch, on the other hand, varies between 150.56 and 336.40 million, depending on the path taken by the sample. When benchmarking with a dataset, the average number of operations will fall between these two numbers. \fig{macs} shows how the confidence threshold affects the average number of multiplication operations.

Choosing a value higher than 0.957 is not meaningful for this model, since the number of operations exceeds that of the Tiny U-Net while yielding worse performance. Conversely, selecting a value of $T$ below 0.895 yields no benefit, as the extra weights of the full network are loaded without improving performance. Therefore, the chosen value for the confidence threshold should lie in the range $\left]0.895, 0.957\right[$. \tab{tab1} shows some values of $T$ and the respective IoU and percentage of inputs of the test dataset that exited at the early branch, as well as the reduction in the number of multiplications when compared to the original Tiny U-Net.

Remarkably, a reduction in the number of multiplications has a positive impact on latency and, more importantly, on energy consumed. This translates into a potential increase in the flight time of the UAV platform.

\begin{table*}[p]
\caption{Impact of the confidence threshold on the model's performance, computational cost, and power}
\begin{center}
\begin{tabular}{c|cc|ccc}
\hline
\mr{3}{\textbf{Confidence} \\ \textbf{threshold,} $\boldsymbol{T}$} & \multicolumn{2}{|c|}{\textbf{Floating Point}} & \multicolumn{3}{|c}{\textbf{Quantized}} \\
& \mr{2}{Mean IoU} & \mr{2}{Exits at early \\ branch (\%)} & \mr{2}{Mean IoU} & \mr{2}{Exits at early \\ branch (\%)} & \mr{2}{Reduction in \\ MACs (\%)}\\
& & & & & \\
\hline
$0.850 \text{ -- } 0.865$ & $0.7611$ & $100.0$ & $0.7629$ & $100.0$ & $42.22$ \\
$0.866 \text{ -- } 0.895$ & $0.7618$ & $98.04$ & $0.7629$ & $100.0$ & $42.22$ \\
\hline
$0.915$ & $0.7619$ & $96.08$ & $0.7643$ & $96.08$ & $39.42$ \\
$0.930$ & $0.7666$ & $78.43$ & $0.7660$ & $88.24$ & $33.83$ \\
$0.940$ & $0.7710$ & $58.82$ & $0.7679$ & $80.39$ & $28.23$ \\
$0.945$ & $0.7746$ & $41.18$ & $0.7698$ & $72.55$ & $22.64$ \\
$0.950$ & $0.7805$ & $23.53$ & $0.7724$ & $60.78$ & $14.25$ \\
$0.957$ & $0.7844$ & $9.80$ & $0.7767$ & $41.18$ & $0.26$  \\
\hline
$0.981 \text{ -- } 0.990$ & $0.7869$ & $0.0$ & $0.7902$ & $0.0$ & $-29.11^{\mathrm{*}}$ \\
\hline
\hline
Ref. \cite{tu} & $0.79$ & - & $0.79$ & - & -  \\
\hline
\multicolumn{5}{l}{$^{\mathrm{*}}$Negative values mean an increase in the number of operations.}
\end{tabular}
\label{tab1}
\end{center}
\end{table*}

\subsection{Dynamic Power Estimation}

The dynamic power consumption of the proposed architecture was estimated using the number of multiply–accumulate (MAC) operations and the power values reported in \cite{tu}. Based on \cite{tu}, the reference network exhibits a measured dynamic power consumption of 203 mW under 425 million MAC operations, corresponding to 4.78$\cdot$10$^{-\text{10}}$W/MAC.

The early-exit branch of the proposed model requires 151 million MAC operations, while the complete forward path (including the EE block) requires 336 million MAC operations. Consequently, their estimated dynamic power consumptions are 72 mW and 160 mW, respectively.

It should be noted that the network \cite{tu} assumes an input dimension of $128\times128$, whereas this work adopts $100\times100$. For consistency, the number of MAC operations of the reference network \cite{tu} at an input size of $100\times100$ is estimated to be 261 million, resulting in an estimated dynamic power consumption of 124 mW. This highlights the early exit contribution in reducing dynamic power consumption.

\subsection{Analysis of the Generated Maps}

\begin{figure*}[p]
\centerline{\includegraphics[width=\linewidth]{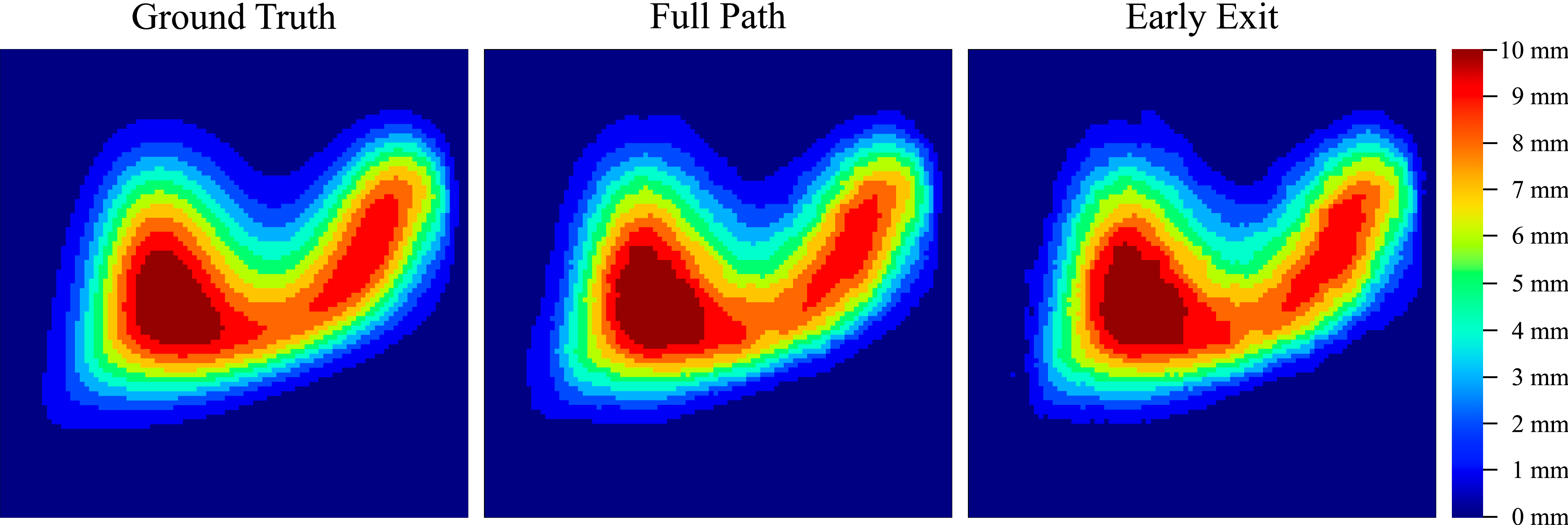}}
\caption{Comparison between the ground truth of an instance of the dataset, along with the estimated maps by the full model and the early exit branch.}
\label{maps}
\end{figure*}

\fig{maps} shows the ground truth thickness map and the estimated maps generated by the EE branch and the full model, respectively, for an instance of the dataset. The borders of each thickness layer are visibly smoother and more coherent in the map generated by the full pass, which is a result of having an extra layer of encoding/decoding. Some noisy pixels can also be seen more frequently in maps generated by the early exit branch, especially when the instance contains oil near the borders.

\begin{figure*}[p]
\centering
\subfloat[1\,mm]{\includegraphics[width=0.48\linewidth]{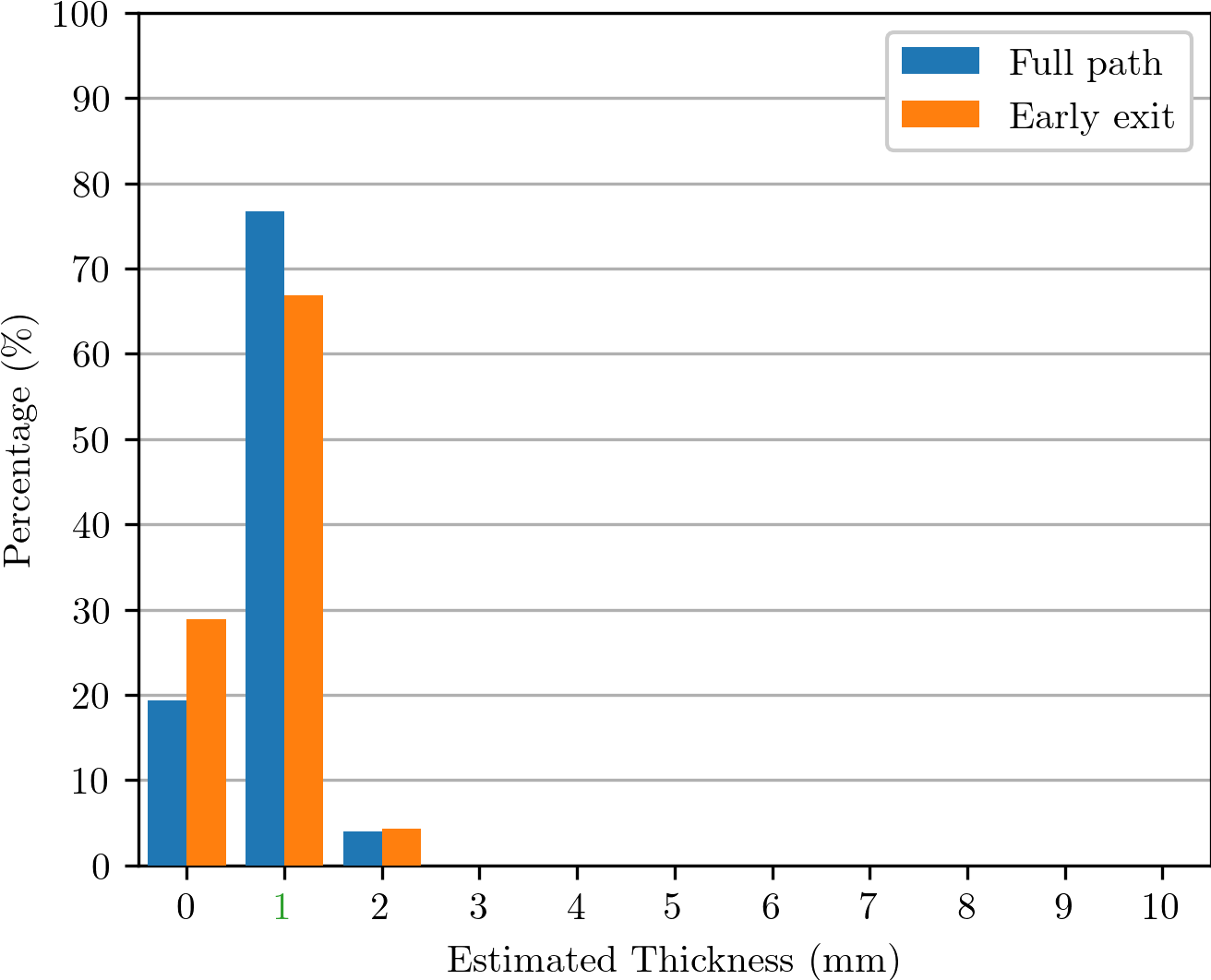}}
\hfill
\subfloat[5\,mm]{\includegraphics[width=0.48\linewidth]{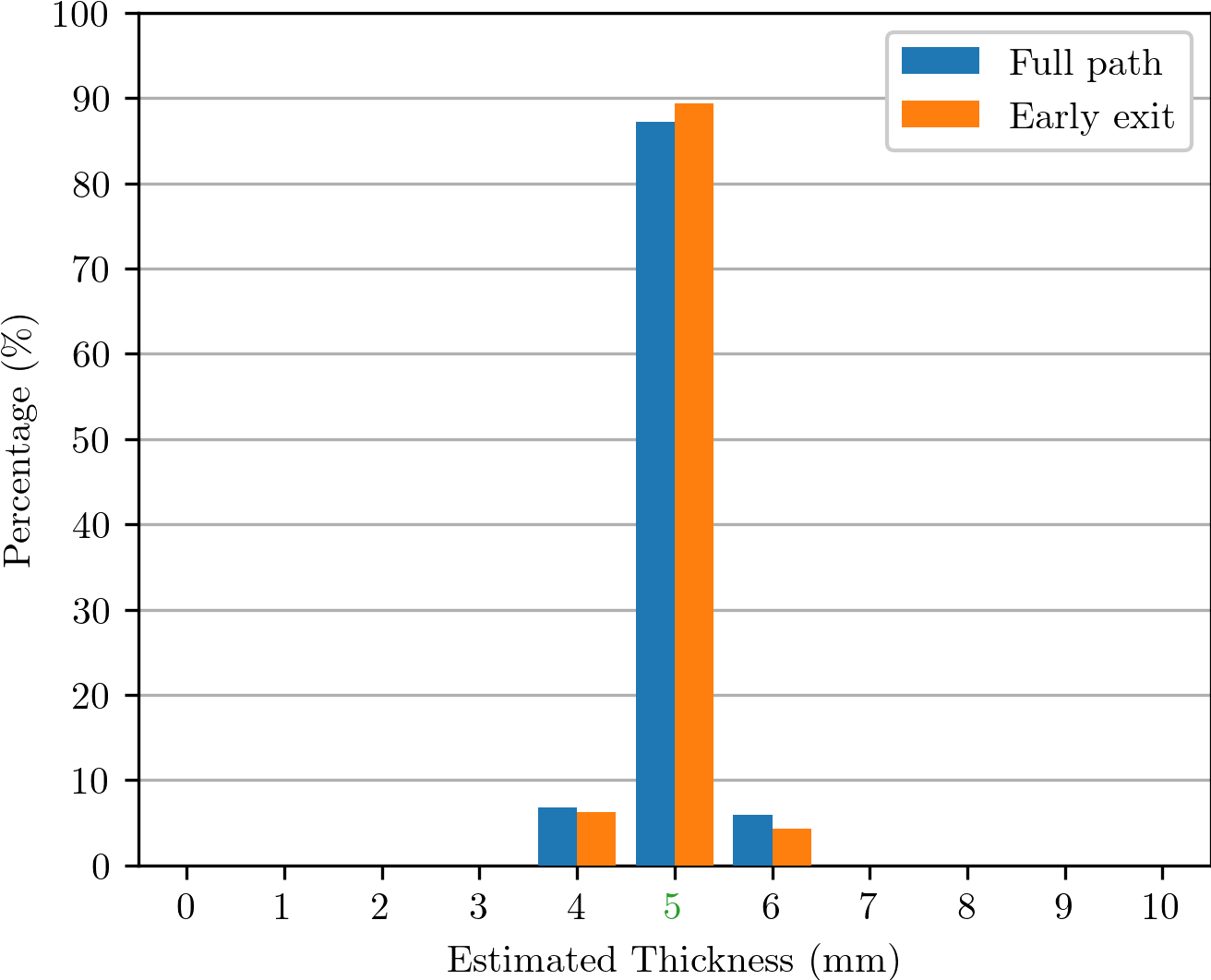}}
\caption{Histograms of estimated thickness for pixels of different true values of a single instance.}
\label{histogram_instance}
\end{figure*}

We should also consider how the different approaches misestimate the thickness values. \fig{histogram_instance} shows four different histograms, considering oil patches of different thicknesses (1 and 5\,mm), and different scopes (single instance/full dataset). We can see that, for both the single instance and for the full dataset, the EE branch cannot differentiate between 0 and 1\,mm as well as the full model. However, the opposite is true when differentiating between 5\,mm and its neighbor values.

\begin{figure}[ht]
\centerline{\includegraphics[width=0.96\linewidth]{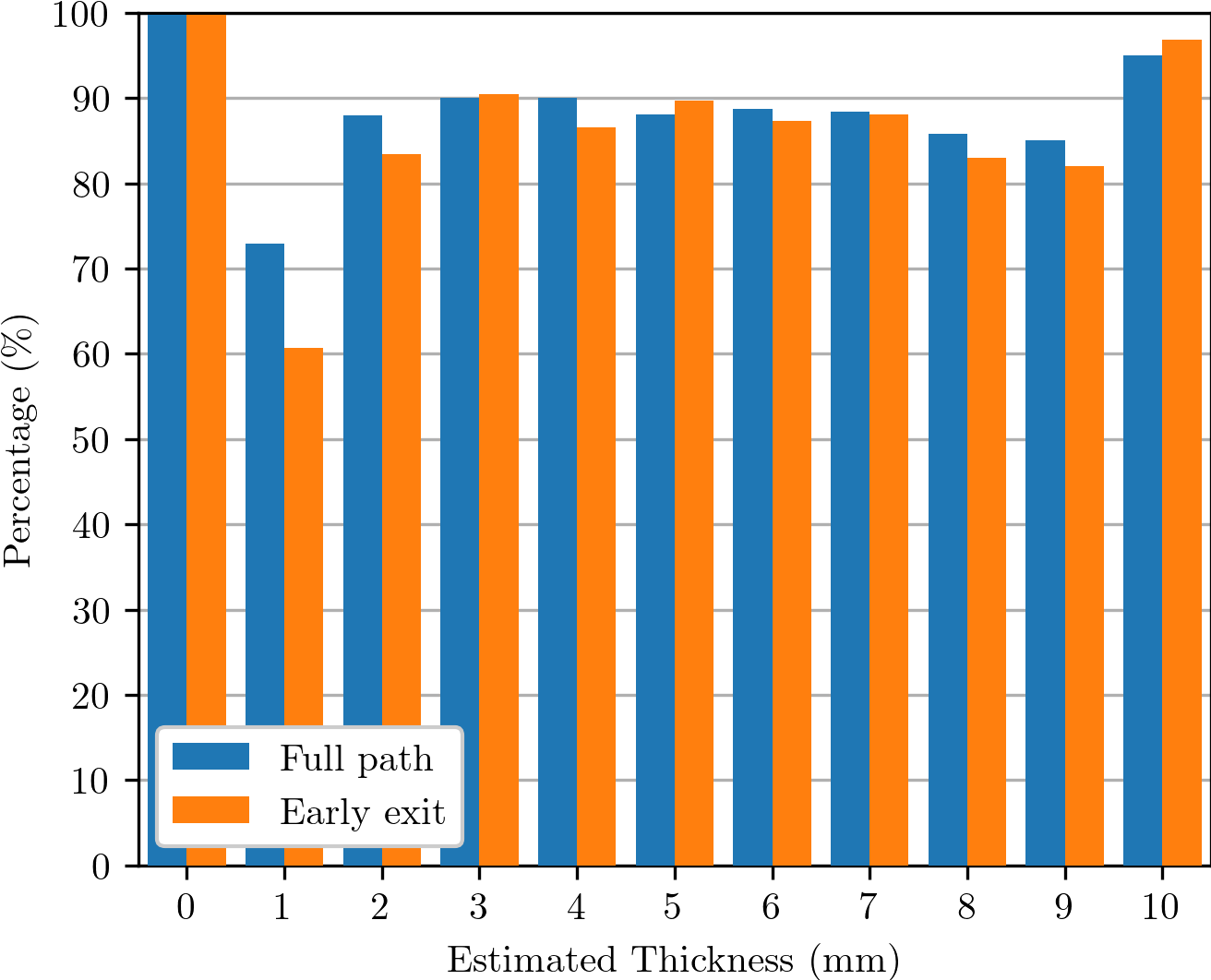}}
\caption{Accuracy of the models for each thickness value across the entire dataset.}
\label{histogram_full}
\end{figure}

\subsection{Limitations and Future Work}
While this work demonstrates the potential of EE mechanisms for efficient UAV-based oil spill monitoring, we acknowledge the following limitations.
Our work builds directly on the very recent work \cite{tu} using the same dataset and model, where all experiments rely on synthetically generated radar data, not fully capturing the complexity and variability of real-world oil spills. This might limit the generalizability to real-world deployments. Additionally, given the Tiny U-Net structure, we explore only a single EE branch configuration with a static confidence threshold $T$.
Furthermore, fully unrolled FPGA architectures limit demonstration of early-exit advantages, as static power consumption dominates the overall energy profile and remains constant irrespective of the executed branch.

Consequently, future work includes validating the design on a more realistic dataset. It also targets an adaptive threshold adjustment policy based on environmental conditions or mission requirements, which could potentially further improve efficiency. Additionally, it should consider an optimized hardware implementation on the FPGA or alternative hardware platforms to more effectively highlight the benefits of early exit mechanisms. Finally, another future research direction is to extend to other environmental monitoring tasks beyond oil spill thickness estimation.
\section{Conclusion}

This work introduces an early exit mechanism in a Tiny U-Net architecture designed for UAV-based radar imaging of oil spills.
After introducing the early exit branch, we were able to maintain similar performance while providing a reduction in the average computational cost of the model. There is, nonetheless, a trade-off between performance and efficiency.
Across the threshold range, IoU varies by up to 2.7\% while multiplications decrease by up to 42\% versus baseline Tiny U-Net. Choosing an intermediary value for the confidence threshold, such as 0.945, reduces the loss in the IoU metric to 2\% and provides a reduction of more than 20\% in the multiplications and estimated power consumption.
Our proposed approach directly targets onboard processing for UAVs. Lowering computation and power extends UAV mission time, enables edge AI deployment for radar-based sensing, and impacts real-time environmental monitoring.

\end{document}